\documentstyle[twocolumn,aps,pra,epsfig]{revtex}
\def\be{\begin{equation}}
\def\e#1{\label{#1}\end{equation}}
\def\bea{\begin{eqnarray}}
\def\ea#1{\label{#1}\end{eqnarray}}

\def\ee{\end{equation}}
\def\eea{\end{eqnarray}}
\def\bem#1{\begin{mathletters}\label{#1}}
\def\eml{\end{mathletters}}

\begin{document}
\title{Collective Excitations of a ``Gravitationally'' Self-Bound Bose Gas}
\author{S. Giovanazzi$^{(a)}$  \and G. Kurizki$^{(a)}$
\and I. E. Mazets$^{(b)}$ \and and S. Stringari$^{(c)}$}
\address{ 
(a) Department of Chemical Physics, Weizmann Institute of Science,
76100 Rehovot, Israel,\\
(b) Ioffe Physico-Tecnical Institute, 194021 St.Petersburg, Russia,\\
(c) Dipartimento di Fisica, Universita di Trento and Istituto 
Nazionale per la Fisica della Materia, I-38050 Povo, Italy.}
\date{\today}
\maketitle

\begin{abstract}
We investigate the collective excitations of an  atomic 
Bose-Einstein condensate in the self-binding regime  produced by 
electromagnetically induced ``gravity'' ($1/r$ attraction).
Analytical expressions for the frequencies of the  
monopole and quadrupole modes are obtained at zero temperature,
using the sum-rule approach,
and compared with the exact results available in the Thomas-Fermi limit.
The low-energy dynamics of such condensates is shown to be dominated 
by the effective ``plasma'' frequency. 
An analog of the Jeans gravitational instability is analyzed.

\vskip 3pt

PACS number(s): 03.75.Fi, 34.20.Cf, 34.80.Qb, 04.40.--b
\end{abstract}

\vskip 6pt

\emph{Introduction} --- 
Recently, a new kind of atomic Bose-Einstein condensates 
has been proposed \cite{odell2000}. 
It has been shown that particular configurations of intense, off-resonant 
electromagnetic fields mimic the gravitational attraction  between atoms
located well within the wavelength of these fields, giving rise to scenarios
very different from the ones  characterizing usual Bose-Einstein 
condensates \cite{s1995,s1999}.
The interatomic potential is of the form $-u/r$, where the gravitational 
coupling $u = Gm^2$ is replaced by 
\be
u = \frac {11   }{60 \pi \epsilon _0^2c} \alpha^2 q^2 I \;.
\label{equ}
\ee
Here the total laser intensity and wavenumber 
are denoted by $I$ and $q$, respectively, and $\alpha$ is the atomic 
polarizability at the  frequency $c q$ \cite{odell2000}.
For a sufficiently strong induced ``gravity'' the BEC becomes  self-bound,
i.e. stable in the absence of an external confining potential 
\cite{giova2001}.
Such a self-bound system may become a laboratory analog of a non-relativistic 
Bose star \cite{ruffini69}, a self-gravitating system that balances
gravity with the ''quantum pressure''  
corresponding to the zero-point kinetic energy fluctuation.
In the self-bound  condensate considered in Ref. \cite{odell2000}
the induced ``gravity'' can be also balanced by the  short-range 
interatomic repulsion.

It is known that the long-wavelength collective excitations of a 
\emph{spatially-uniform} system of both fermions (e.g., the electron gas) 
and bosons \cite{foldy61}, interacting via a \emph{repulsive}
potential of the form $u/r$, are characterized by a 
spectral gap, fixed by the ``plasma'' frequency 
\be
\omega_p^2 = 4 \pi u n / m \;, 
\label{omegap}
\ee
where $n$ is the particle density.
In this paper we show that also in the case 
of a \emph{self-bound} atomic BEC,
interacting via an \emph{attractive} $-u/r$ potential, the collective 
oscillations are characterized by an effective ``plasma'' frequency
$\omega_p$  of the form (\ref{omegap}), 
with a spatially-averaged density $n$
and the positive coupling constant $u$ [Eq.\ (\ref{equ})]. 
Its correspondence with the fundamental oscillation frequency
of white dwarfs and neutron stars \cite{shapiro} is pointed out
and the analog of the Jeans gravitational instability  is explicitly 
analyzed.

\emph{Generalized Gross-Pitaevskii equation} --- 
The Gross-Pitaevskii equation \cite{Pitaevskii61}
for the order parameter $\Psi ({\bf r},t)$ can be obtained starting from
\be
i \hbar  \frac{\partial \Psi }{\partial t} =
\frac{\delta }{\delta \Psi^* } H_{\mathrm{tot}}\;.
\label{ggp}
\ee
Here the total mean-field energy functional 
$H_{\mathrm{tot}} = H_{\mathrm{kin}} + H_{\mathrm{ho}}+
H_{\mathrm{grav}} + H_{\mathrm{s}}$ consists of:
(a) the kinetic energy,  
$H_{\mathrm{kin}}=\int \,d{\bf r} (\hbar^2/2m)|{\bf \nabla}\Psi|^2$;
(b) the harmonic-trap energy,  $H_{\mathrm{ho}}=\int 
V_{\mathrm{ho}}\,n \,d{\bf r} $, where 
$n = |\Psi({\bf r},t)|^{2}$ is the atomic density
and $V_{\mathrm{ho}} = m\omega_{0}^2 r^2 / 2$ is an isotropic harmonic 
external potential;
(c) the mean-field energy due to the electromagnetically induced ``gravity'', 
$ H_{\mathrm{grav}} =   (1/2) \int n \, \Phi \,d{\bf r}$,
where $\Phi$ is the ''gravitational potential'' 
solution of the Poisson equation 
\be
{\bf \nabla}^2\Phi=4\pi u n \;,
\label{poisson}
\ee 
with the boundary condition $\Phi=0$ for $|r| \rightarrow \infty$;
(d) the mean-field contribution due to the short-range scattering, 
$H_{\mathrm{s}} = (g/2) \int n^{2} \;d{\bf r} $,
where $g= 4 \pi a \hbar^{2} / m$, 
$a$ being the s-wave scattering length.

By expressing the complex BEC order parameter through the condensate density 
$n$ and phase $\phi$ as  $\Psi = \sqrt{n} \exp(i\phi )$,
we can write Eq. (\ref{ggp}) as a coupled set of 
collisionless hydrodynamic equations:
\begin{equation}
\frac {\partial n}{\partial t}+ {\bf \nabla} (n{\bf v})=0 ,  
\label{continuity}
\end{equation}
\be
m\frac {\partial {\bf v}}{\partial t}
+ {\bf \nabla} \left( \frac {m v^2}2
-\frac {\hbar ^2 \nabla ^2\sqrt{n}}{2m\sqrt{n}}+
g n + V_{\mathrm{ho}} + \Phi \right)   = 0 \;.
\label{twoeq}
\ee
The last equation establishes the irrotational property of the superfluid 
velocity field defined by ${\bf v}=\hbar  {\bf \nabla} \phi /m $. 
Similar equations have been derived in the case of harmonically trapped 
Bose-Einstein condensates interacting via zero-range forces
\cite{stringari96}.

\emph{Jeans-like instability} ---
In the absence of external potentials, Eqs. (\ref{continuity}) and 
(\ref{twoeq}) admit a stationary solution with constant density $n$. 
Such a solution is, however, dynamically unstable, as can be seen 
directly from the  Bogoliubov dispersion relation \cite{bogo47}
corresponding to (\ref{continuity}) and (\ref{twoeq}):
\be
\omega^2_k = \frac{\hbar^2 k^4}{4m^2} +  c_s^2 k^2 - \omega_p^2
\label{bogo}\;.
\ee 
In (\ref{bogo}) $c_s=\sqrt{g n /m}$ is the Bogoliubov speed of sound
and the plasma frequency $\omega_p$ is given by (\ref{omegap}). 
The occurrence of imaginary frequencies $\omega_k$ for small values of 
$k$ reveals the existence of an instability,
which resembles the Jeans gravitational instability \cite{ruffini69,astro}.
Assuming $\omega_p \ll m c_s^2$, the minimal wavelength ensuring a real value
for the frequency (\ref{bogo}) is given by 
\be
\lambda_J = 2 \pi c_s / \omega_p = 2 \pi (\hbar^2 a / m u )^{1/2}\;.
\label{jeanslength}
\ee
In a ``self-gravitating''  BEC, both  $\omega_p$ and $c_s$ 
scale as $n^{1/2}$, so that the critical wavelength
$\lambda_J$ is {\emph independent of the density},
as explicitly revealed by (\ref{jeanslength}).

Besides these unstable solutions, Eqs. (\ref{continuity})
and (\ref{twoeq}) 
admit a stable self-bound solution for any finite number N of atoms. 
In this case long-wavelength fluctuations, 
which can trigger  the Jeans instability, are excluded by 
the finite size of the system.
Remarkably, the ground-state density 
of a self-bound Bose cloud, 
in the limit in which the kinetic energy can be neglected 
(Thomas-Fermi limit for Bose gases),
has the analytic profile \cite{odell2000}
\be
n_0({\bf r})=\frac N{4 R_0^2}\frac {\sin (\pi r/R_0)}r 
\theta (R_0-r)  = 0,  \label{n0}  
\ee
where the \emph {radius}  $ R_0 = \lambda_J / 2$ is 
\emph {fixed by the Jeans wavelength}
and $\theta $ is the Heaviside function.

\emph{Sum rule approach} --- 
The quadrupole and monopole excitation frequencies at zero temperature
can be estimated using a sum-rule approach \cite{stringari96,report}.
The calculation of the oscillation frequencies 
$\omega$ is based on the ratio 
\begin{eqnarray}
\hbar^2\omega^2 = {m_3}/{m_1} 
\label{omega31}
\end{eqnarray}
between the third moment (cubic-energy weighted) 
and the first moment (energy-weighted)
of the  dynamic structure factor: 
$m_p=\sum_n (\hbar\omega_{n0})^p \mid<0\mid F\mid n>\mid^2$
$(p=1,3)$.
Here $\hbar\omega_{n0}$ is the excitation energy of the state $|n>$ 
and $F$ is an excitation operator chosen
similarly to \cite{stringari96}:
$F = \sum_{i=1}^N r^2_i$ or  $F = \sum_{i=1}^N r^2_i Y_{2m} $,
for the  monopole or quadrupole operators, respectively.
The moments $m_1$ and $m_3$ can be reduced to the form of 
commutators involving the Hamiltonian of the system 
and evaluated in terms of ground-state expectation values.
One then obtains the following results:
\bea
\omega^2_M &=&  \frac
{ 4 H_{\mathrm{kin}} + 4 H_{\mathrm{ho}} + H_{\mathrm{grav}}
+ 9 H_{\mathrm{s}} } 
{ N m \langle r^2 \rangle }\nonumber\\
&=& \frac 
{ 2 H_{\mathrm{kin}} + 6 H_{\mathrm{ho}} + 6 H_{\mathrm{s}} } 
{ N m \langle r^2 \rangle }\;, 
\label{monosr} \\
\omega^2_Q &=&  \frac 
{ 4 H_{\mathrm{kin}} + 4 H_{\mathrm{ho}} - 4/5\; H_{\mathrm{grav}} } 
{ N m \langle r^2 \rangle } \;,
\label{quadsr}
\eea
where, in deriving the second equation (\ref{monosr}),
we have used the virial identity
\be
2 H_{\mathrm{kin}} - 2 H_{\mathrm{ho}} + H_{\mathrm{grav}}
+ 3 H_{\mathrm{s}} = 0\;.
\label{viriale}
\ee
The frequencies $\omega_M$ and  $\omega_Q$, given by
(\ref{monosr}) and  (\ref{quadsr}), 
are always real for positive values of the scattering length.
If the expectation values of Eqs. (\ref{monosr}) and  (\ref{quadsr})
are evaluated using the \emph{exact} ground state,
they provide a \emph{rigorous upper bound} to the 
lowest excitation frequencies for a given multipole.
This approach has been shown to provide an excellent description of the 
crossover between the non-interacting and the Thomas-Fermi regimes 
in the case of condensates interacting via zero-range forces
\cite{s1999}.

A useful analytical estimate 
for the expectation values of Eqs. (\ref{monosr}) and  (\ref{quadsr})
can be obtained  
by replacing the exact ground state with the \emph{gaussian wavefunction}
$ \Psi = N^{\frac 12}
\exp \left(-r^2 / 2 w^2  )  \right) / \pi^{{3\over4}} w ^{{3\over2}} $.
Here $w$ is a variational parameter that can be calculated by minimizing the 
total energy $H_{\mathrm{tot}}$ and is related to the expectation value
$\langle r^2 \rangle= (3/2) \; w^2 $.
The various contributions to the energy $H_{\mathrm{tot}}$ then take the form:
$H_{\mathrm{kin}}$ $=$ $ 3 N \hbar^2 / 4 m w^2$,
$H_{\mathrm{ho}}$ $=$ $  N m \omega_0^2 w^2/4 $,
$H_{\mathrm{s}} $ $=$ $ a \hbar^2 N^2$ $/$ $ \sqrt{2\pi} m w^3 $ 
and $ H_{\mathrm{grav}} = - (1/\sqrt{2\pi}) u N^2 / w $.
The gaussian ansatz %(\ref{eq:ansatz}) 
allows us to describe analytically the various regimes in the 
"phase diagram" of the problem \cite{odell2000}.
It is noteworthy that the monopole and quadrupole frequencies obtained by
the gaussian ansatz coincide with the ones derivable from a 
time-dependent variational calculation in a gaussian basis \cite{perezg96}.

\emph{Low-energy excitations in a self-bound BEC} --- 
In the following we calculate the monopole and quadrupole
excitation frequencies for a self-bound BEC from Eqs. (\ref{monosr}) 
and  (\ref{quadsr}), using, for the ground state, either the {\emph exact} 
solution of the Gross-Pitaevskii equation or the gaussian approximation.
In these calculations we are interested in the effects of  s-wave scattering,
measured by the dimensionless parameter 
\be
s = m u N^2 a/\hbar^2 \;.
\ee

Upon neglecting the role of the external trap,
the width of the ground state, within the gaussian ansatz, is given by
$w = (3/2) \sqrt{\pi/2}   \left(1 + \sqrt{1 + (8/3\pi) s}\right) \Lambda$, 
where $\Lambda = \hbar^2 / m u N $ is the effective "gravitational"
radius which fixes the size of the condensate in the absence of $s$-wave
interaction ($a=0$).
This expression for $w$ allows for the analytical evaluation of the 
monopole and quadrupole frequencies 
(\ref{monosr}) and (\ref{quadsr}) that are plotted in Fig 1
in units of the "gravitational" frequency $m u^2 N^2 / \hbar^3$.
Comparison with the results obtained using the exact ground state
shows that the gaussian ansatz provides indeed a good 
approximation, for both positive and negative values 
of $s$, provided one is far from the critical value $s_c\approx -1$, where 
the condensate exhibits collapse \cite{odell2000}.

The order of magnitude of $\omega_M$ and $\omega_Q$ is given by 
the ``plasma'' frequency $\omega_p$, 
\emph{irrespective} of the scattering parameter $s$.
This can be simply seen
by using the virial identity (\ref{viriale}).
Assuming $s>0$ and setting $ H_{\mathrm{ho}} = 0$ 
(absence of external trapping), we derive the inequalities 
$0 < 2 H_{\mathrm{kin}} < - H_{\mathrm{grav}}$
and, hence, from (\ref{monosr}) and (\ref{quadsr}), we obtain
$1 < \omega^2_M/\bar{\omega}^2 <2$ 
and $ (4/5) < \omega^2_Q/\bar{\omega}^2 <(14/5)$.
Here $\bar{\omega}^2 \equiv - H_{\mathrm{grav}} / N m \langle r^2 \rangle$ 
turns out to be always proportional to the plasma frequency
$\omega_p^2 =4 \pi u n / m $.
%$ 0 \le \omega^2_Q/\bar{\omega}^2 + 2\omega^2_M/\bar{\omega}^2 <(24/5)$.
Similarly, assuming $s_c < s \le 0$, we obtain
$0 < \omega^2_M/\bar{\omega}^2 \le 1$ 
and $ (14/5) \le \omega^2_Q/\bar{\omega}^2 < (24/5)$.
The proximity of the monopole frequency $\omega_M$ to 
the ``plasma'' frequency $\omega_p$ 
breaks down upon approaching the critical value $s_c\approx -1$,
where $\omega_M^2 \rightarrow 0 $ 
(whereas $\omega_Q^2 \rightarrow (24/5)\, \bar{\omega}^2$).

%%%%%%%%%%%%%%%%%  F I G U R E %%%%%%%%%%%%%%%%%%%%%%
\begin{figure}[h]
\vspace{-0.3in}
\begin{center}
%\centerline{\psfig{figure=../fig/special.2.eps,height=3.15in}} 
\centerline{\psfig{figure=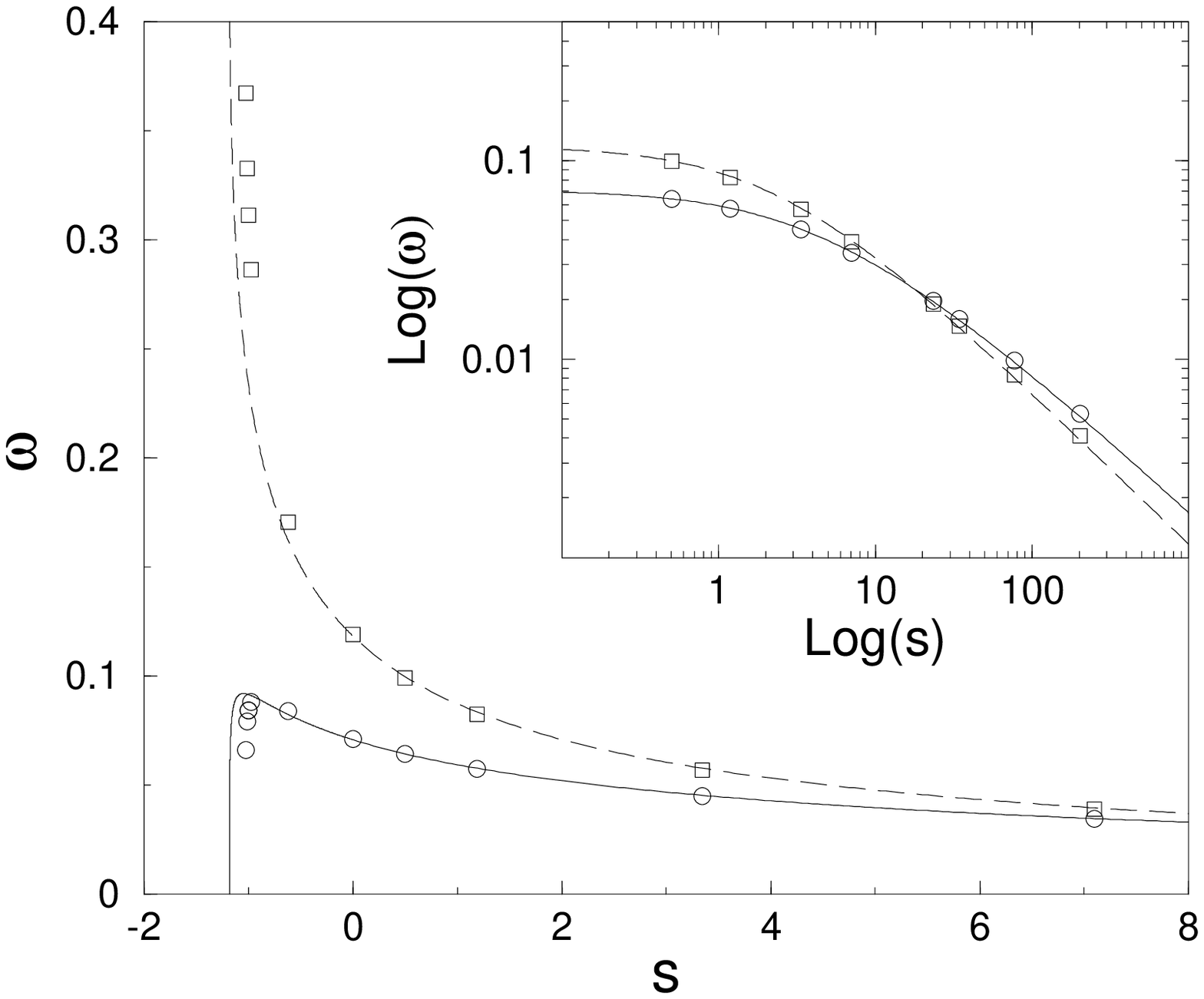,height=3.15in}} 
\end{center}
\vspace{-0.5in}
%%%%%%%%%%%%%%%%%  F I G U R E %%%%%%%%%%%%%%%%%%%%%%
\begin{caption}
{Monopole and quadrupole frequencies of a self-bound condensate (in 
$m u^2 N^2 / \hbar^3 $ units) versus the scattering parameter 
$s= m u N^2 a / \hbar^2 $, for both positive and negative values of $a$. 
Circles: monopole frequencies; squares: quadrupole frequencies.  
The solid  and long-dashed lines  correspond to the gaussian approximation.
Inset: the corresponding Log-Log plot for $a>0$.}
\end{caption}
\label{fig:timegraph}
\end{figure}
%%%%%%%%%%%%%%%%%%%%%%%%%%%%%%%%%%%%%%%%%%%%%%%%%%%%%%

The limits of large and small values of the scattering parameter $s$ 
(denoted as the TF-G and G self-bound regimes, respectively 
\cite{odell2000}) will now be discussed separately:

(a) \emph{TF-G-regime} --- 
In the TF-G-regime (self-bound Thomas-Fermi ``gravitational'' regime) 
\cite{odell2000} ``gravity'' is balanced by the s-wave interaction
 (large-$s$ limit), 
while both the kinetic  energy $H_{\mathrm{kin}}$ ($\propto\hbar^2$),
and the external trap potential are negligible.
In this regime the monopole and quadrupole frequencies 
in (\ref{monosr}) and (\ref{quadsr})
are given, using the ground state profile (\ref{n0}), by
$\omega_M  =  0.6225 \, \omega_p$, $\omega_Q  =  0.3928 \, \omega_p$,
where 
\be
\omega_p =  0.5642 \,
m^{\frac14} u^{\frac54} N^{\frac12} \hbar^{-\frac32} a^{-\frac34}
\label{plasmaTFG}
\ee
is the plasma frequency evaluated at the central density. The ratio  
$\omega_M / \omega_Q$ turns out to be equal to $ \sqrt {5 / 2} $.
Remarkably, this is the same result holding for a \emph{harmonically} 
trapped BEC  in the Thomas-Fermi limit in the absence
of the  $1/r$  interaction \cite{stringari96}.
In the TF-G regime, the chemical potential 
$ \mu = (H_{\mathrm{kin}} + H_{\mathrm{ho}}
+ 2 H_{\mathrm{grav}} + 2 H_{\mathrm{s}} ) / N$
has the value $-\mu = m c_s^2 >> \hbar \omega_p$.

It is interesting to note that  in the Thomas-Fermi limit
the sum-rule expression (\ref{monosr}) for the monopole frequency reduces 
to $\omega^2_M = 2 | H_{\mathrm{grav}} | / N m \langle r^2 \rangle$ and 
coincides with the value derived in politropic stars
using classical hydrodynamics \cite{shapiro}.

In the TF-G regime the static linear response for the monopole 
operator $F = \sum_{i=1}^N r^2_i$ 
can be calculated analytically.
This allows for an alternative way to estimate the monopole frequency 
through the ratio $\hbar^2\omega^2 = m_1/m_{-1}$, where 
the inverse-energy weighted moment $m_{-1 }$ is
directly related to the static response \cite{report}.
In the limit of a small harmonic perturbation $\lambda r^2$,
the induced density fluctuation takes the analytic form
\be
\delta n = \frac{\lambda}{2\pi u}
\left( \frac{\sin(\pi r/R_0)}{r / \pi R_0} - 3 \right) \theta (R_0-r)
\;,
\ee
and the inverse-energy weighted moment becomes 
$m_{-1} = - \delta \langle r^2 \rangle /  2 \lambda =
 (6\pi^3 - (2/5)\pi^5) \, m^{-\frac52} u^{-\frac72} \hbar^{5} a^{\frac52}$. 
In terms of the plasma frequency (\ref{plasmaTFG}), we finally
find $\omega_M = 0.6181 \, \omega_p$, 
a value slightly smaller  than the  one derived from 
the ratio $m_3/m_1$.

(b) \emph{G-regime} --- 
We turn now to the asymptotic G-regime 
(self-bound ``gravitational'' regime) \cite{odell2000},
in which the s-wave interaction and the external trap potential 
are negligible, and ``gravity'' is balanced by the kinetic  energy 
$H_{\mathrm{kin}}$.
In this case the monopole and quadrupole frequencies (\ref{monosr}), 
(\ref{quadsr}) are given by $\omega_M  =   0.3022 \, \omega_p$,
$\omega_Q  =   0.5057 \, \omega_p$, respectively, where 
\be
\omega_p =  0.2351 \, m u^2 N^2 / \hbar^3 
\label{plasmaG}
\ee
is the plasma frequency (\ref{omegap}) evaluated at the central 
density in the same regime.
The ratio between these frequencies is  now
$ \omega_M / \omega_Q  = \sqrt {5 / 14} < 1$.
The chemical potential $ \mu = (H_{\mathrm{kin}} + H_{\mathrm{ho}}
+ 2 H_{\mathrm{grav}} + 2 H_{\mathrm{s}} ) / N$ in this regime has the value 
$ \mu = - 0.6922 \, \hbar \omega_p$.
This shows that both the monopole and quadrupole oscillation frequencies 
are smaller than $ - \mu/\hbar$. 
Consequently, these oscillations  are \emph{stable} against the ejection 
of atoms from the  condensate.

Also in the G-regime the  evaluation of the static monopole response 
can be used to estimate the monopole frequency 
through the ratio $\hbar^2\omega^2 = m_1/m_{-1}$. The calculation yields
$\omega_M  =  0.2928  \, \omega_p$.
Similarly to the the TF-G regime the monopole frequencies evaluated 
through the ratios $m_3/m_1$ and $m_1/m_{-1}$ are remarkably close, 
confirming the validity of the method.

\emph{Collisionless hydrodynamics} ---
In the Thomas-Fermi (TF-G) regime, 
\emph {exact results} for the oscillation frequencies are obtainable
by solving explicitly the equations of collisionless hydrodynamics
(\ref{continuity}), (\ref{twoeq}).
These results allow us to improve the sum-rule estimate, which in general 
provides only an upper bound.
The equations of motion (\ref{continuity}) and (\ref{twoeq}) can  
be simplified in this regime, since the quantum pressure term, 
proportional to $\hbar^2$, can be safely ignored.
After linearization they yield the equation of motion for the density 
fluctuations $\delta n$ 
\be
\frac{\partial^2}{\partial t^2} \delta n
= {\bf \nabla}
 \left[ c_s^2 {\bf \nabla}\left( \delta n + \delta \Phi/g \right) \right] \;,
\label{linear}
\ee
where the ``local'' sound velocity $c_s(r)=\sqrt{g n(r)/m}$
is calculated at the equilibrium  density (\ref{n0}).
The time dependence of $\delta n$ is chosen in the form of $\exp (-i\omega t)$,
and  (\ref{linear}) is transformed, with the help of  (\ref{poisson}), 
into
\be
\tilde \omega ^2\tilde \nabla ^2 \xi +\tilde \nabla \left[ 
\frac {\sin (2\pi\tilde{r})}{2\pi \tilde r}\tilde \nabla 
\left( \frac{\tilde \nabla ^2\xi}{4\pi^2} +\xi \right)  \right]  =0\;,
\label{eqxi} 
\ee
where we have introduced the ''gravitational'' fluctuation variable 
$\xi=\delta \Phi/g$ and used the  dimensionless coordinate 
$\tilde{\bf r} = {\bf r}/\lambda_J$ and frequency 
$\tilde{\omega }=\omega/\omega_p$, with $\omega_p$ given by (\ref{plasmaTFG}).

To solve this equation, we set $\xi =\xi _l (\tilde{r})Y_{lm}
(\vartheta, \varphi )$, where $Y_{lm}$ are the usual  spherical harmonics. 
The boundary conditions can be derived from the tailoring of the inner 
and outer solutions on the condensate surface, i.e. at $\tilde{r}=\pi$
\cite{mazets.arc}. 
The conservation of the total number $N$ of atoms implies 
 that the radial derivative of the potential perturbation vanishes at  
$\tilde r \le \pi $ for the monopole mode. For the modes 
with $l=1,2,3,\, ...$ the function  $\xi _l$ instead decreases as 
$r^{-(l+1)}$. Hence, the boundary conditions are 
$(d\xi _0 / d\tilde r ) \vert _{\tilde r=\pi }=0$ 
for $l =0$ and $ (d\xi_l / d\tilde r) \vert _{\tilde r=\pi }
=-(l+1) \xi _l / \pi$  for  $l>0$ \cite{mazets.arc}.

The numerical solutions of (\ref{eqxi}) for the monopole and quadrupole 
frequencies yield $\omega_M=0.6168\,\omega_p$, $\omega_Q=0.392\,\omega_p$
with $\omega_p$ as in (\ref{plasmaTFG}).
Hence, the sum-rule estimates given above, for these frequencies
turn out to be \emph{very close} (within $1\%$) to the exact numerical values.
Furthermore,  they satisfy  the  inequalities 
predicted by the general formalism of sum rules \cite{report}.

\emph{Conclusions} ---
In this paper we have mainly dealt with the oscillations of a 
``gravitationally'' self-bound bosonic cloud,
in the framework of the (generalized) Gross-Pitaevskii approach.
The Jeans wavelength (\ref{jeanslength}), which is the shortest wavelength 
to ensure stability, has been shown to be the diameter of the self-bound
cloud in the Thomas-Fermi limit [Eq. (\ref{n0})].
We have derived here simple, analytical expressions for the 
frequencies of the lowest monopole and quadrupole modes, 
which are the most relevant collective excitations of such a system,
through a sum-rule approach.
We have also numerically calculated the eigenfrequencies of the 
linearized generalized Gross-Pitaevskii equation 
in the Thomas-Fermi limit of a self-bound BEC and demonstrated 
that the results are very close to the sum-rule values.

The studied excitations substantially differ from those
of harmonically trapped gases interacting via zero-range forces
\cite{stringari96,perezg96}.
For harmonically-trapped bosons, all the low-energy excitations
(monopole, dipole, quadrupole etc.) are fixed by the oscillator frequency
$\omega_0$ of the confining trap.
By contrast, in the self-bound regimes, 
only the frequency of the lowest dipole mode,
associated with the center of mass oscillations, 
is fixed by the external trap frequency $\omega_0$,
whereas all the other low-energy excitations are dominated by 
the effective ``plasma'' frequency $\omega_p$ (\ref{omegap}).

Since the repulsive short-range (s-wave) force can drastically change the 
size of the self-bound atomic cloud \cite{odell2000}, the explicit values 
of $\omega_p$ and the corresponding monopole and quadrupole frequencies 
are determined by the central density  and, consequently, by the regime 
considered [Eqs.\ (\ref{plasmaG}) and (\ref{plasmaTFG})].
The dependence of the low-frequency collective modes on the plasma frequency 
resembles the oscillations of compact stars, such as white dwarfs and 
neutron stars, where the frequency scale $\omega_p^2 \sim 4 \pi G \rho$ 
provides the universal relation between the oscillation timescale
and the mass density $\rho$ \cite{shapiro}.

This work was supported by the German-Israeli Foundation (GIF),  
the Russian Foundation for Basic Research 
(project 99--02--17076) and the State Program ''Universities of 
Russia'' (project 015.01.01.04).

\end{document}